\documentclass[10pt, conference, compsocconf]{IEEEtran}

\usepackage{fancyhdr}
\fancypagestyle{IEEEtitlepagestyle}{
  \fancyhf{}
  \fancyfoot[C]{\sffamily\footnotesize\thepage}
  \fancyfoot[L]{\sffamily\scriptsize Copyright held by the owner/author(s). Presented at\\
3rd International Workshop on Overlay Architectures for FPGAs (OLAF2017)
\\Monterey, CA, USA, Feb. 22, 2017.}

}
\usepackage[T1]{fontenc}
\newcommand{\quotes}[1]{``#1''}
\usepackage{enumerate}
\usepackage{floatrow}
\usepackage{algpseudocode,algorithm,algorithmicx}

\algrenewcommand\algorithmicrequire{\textbf{Precondition:}}
\algrenewcommand\algorithmicensure{\textbf{Postcondition:}}

\ifCLASSINFOpdf
  \usepackage[pdftex]{graphicx}
  \graphicspath{{./img/}}

\else
\fi

\usepackage[cmex10]{amsmath}
\usepackage{subcaption}

\usepackage{url}

\usepackage[color=green!40,textsize=footnotesize,shadow]{todonotes}

\begin{document}
\bstctlcite{IEEEexample:BSTcontrol}

%
\title{\vspace*{-0.40cm}Pixie: A heterogeneous Virtual Coarse-Grained Reconfigurable Array for high performance image processing applications\vspace*{-0.5cm}}


\author{\IEEEauthorblockN{Amit Kulkarni and Dirk Stroobandt}
\IEEEauthorblockA{ELIS department, Computer Systems Lab\\
Ghent University\\
iGent, Technologiepark
Zwijnaarde 15, B-9052 Ghent - Belgium.\\
Email:$\{$Amit.Kulkarni, Dirk.Stroobandt$\}$@UGent.be}
				\and

\IEEEauthorblockN{Andre Werner, Florian Fricke, Michael Huebner}
\IEEEauthorblockA{Chair for Embedded Systems for Information Technology,\\ Ruhr University of Bochum,\\
44780 Bochum - Germany\\
Email:\{Andre.Werner-w2m, Florian.Fricke,\\ Michael.Huebner\}@rub.de}
}


%


\maketitle
\vspace*{-0.8cm}
\vspace*{-0.25cm}\begin{abstract} 
Coarse-Grained Reconfigurable Arrays (CGRAs) enable ease of programmability and result in low development costs. They enable the ease of use specifically in reconfigurable computing applications. 
The smaller cost of compilation and reduced reconfiguration overhead enables them to become attractive platforms for accelerating high-performance computing applications such as image processing. The CGRAs are ASICs and therefore, expensive to produce. However, Field Programmable Gate Arrays (FPGAs) are relatively cheaper for low volume products but they are not so easily programmable. We combine best of both worlds by implementing a Virtual Coarse-Grained Reconfigurable Array (VCGRA) on FPGA. VCGRAs are a trade off between FPGA with large routing overheads and ASICs.
In this perspective we present a novel heterogeneous Virtual Coarse-Grained Reconfigurable Array (VCGRA) called ``Pixie'' which is suitable for implementing high performance image processing applications. 
The proposed VCGRA contains generic processing elements and virtual channels that are described using the Hardware Description Language VHDL. 
Both elements have been optimized by using the parameterized configuration tool flow and result in a resource reduction of 24\% for each processing elements and 82\% for each virtual channels respectively. 
\end{abstract}

%


%
\IEEEpeerreviewmaketitle

\section{Introduction} 
Offloading complex computational tasks of a high performance computing application from a general purpose processor to dedicated hardware is an interesting research that provides an opportunity for efficient design of hardware accelerators. 
Field Programmable Gate Arrays (FPGAs) are very attractive platforms for developing such accelerators.
Thus, FPGAs are used as auxiliary hardware for data intensive computing processing units such as DSPs or GPUs. 

The costs for designing applications on FPGAs is much higher than the cost of developing GPU or DSP implementations, resulting in a wide implementation gap between the application description and its implementation on FPGA. 
The effort (time needed for the algorithms for synthesis, mapping, placement and routing) required during the processing of the high level or hardware description code is the major part of the design costs, resulting in slow design cycles compared to GPUs and DSPs. 
In order to bridge the gap between the high level application description and the FPGA implementation VCGRAs can be used. 
A number of authors has proposed VCGRAs that are predominantly dependent on the hard coded DSP block primitives that are available in modern FPGAs~\cite{deco}~\cite{dsp_overlay}~\cite{esit}. 

The programming model for Virtual Coarse-Grained Reconfigurable Arrays (VCGRAs) allows the user to write the code at a higher abstraction level without worrying about every low level detail of the architecture. 
This reduces the compilation time by several orders of magnitude compared to the compilation time for the Fine-Grained FPGAs thus VCGRAs are overlay architectures that help to curb the development costs. 
\vspace*{-0.1cm}
\subsection*{VCGRA architecture}
The VCGRA consists of coarse-grained element (called processing element, PE) groups connected using virtual connection blocks and switch blocks forming a communication network (inter-connect). 
The processing elements are powerful and more complex than a LUT and are defined at a higher abstraction level. 
The complexity of the processing elements can range from a simple ALU to a fully capable RISC processor. 
Each PE has a settings register used to configure the function of the PE. 
With the proper connection settings (configured in the settings register of the VSB - Virtual Switch Block), every application that uses these PEs can be implemented. 
The settings registers are updated using a dedicated bus that enables us to reconfigure the settings of the PEs and VSBs.



The VCGRA grid can be efficiently implemented using the parameterized configuration tool flow~\cite{amit_5}. The parameterized configuration is an optimization technique used for implementing a parameterized application on an FPGA. The application is said to be parameterized when some of its inputs, called parameters, are infrequently changing compared to the other inputs. Instead of implementing these parameter inputs as regular inputs, in the parameterized configuration approach these inputs are implemented as constants and the design is optimized for these constants. When the parameter values change, the design is re-optimized for the new constant values by reconfiguring the FPGA.

The PEs are implemented efficiently using a constant propagation approach and the intra-connects of each PE along with the VCGRA interconnection network are mapped on to the parameterized physical Switch Blocks (SBs) and Connection Blocks (CBs) implemented using tunable connections (TCONs)~\cite{elias_tpar}.

In this paper we propose a general purpose heterogeneous VCGRA grid suitable to implement digital image processing filters and other math operations. 
The PEs of the proposed VCGRA grid are identical to each other. 
However, the PEs can be configured to perform different arithmetic operations such as Add, Sub, Mul, Div, etc. and hence they emulate heterogeneity in their functions. 
The rest of the paper is organized as follows: 
Section~\ref{sec:soa} presents the state of the art. 
The proposed heterogeneous VCGRA grid is described in Section~\ref{sec:vcgra_grid}. Section~\ref{sec:application} presents the Sobel edge detection filter implemented on the proposed VCGRA. 
The results are presented in Section~\ref{sec:results} followed by the concluding remarks in Section~\ref{sec:conclusion}.

\vspace*{-0.4cm}
\section{State of the Art}\label{sec:soa}
\begin{figure}
\centering
\includegraphics[width=7cm]{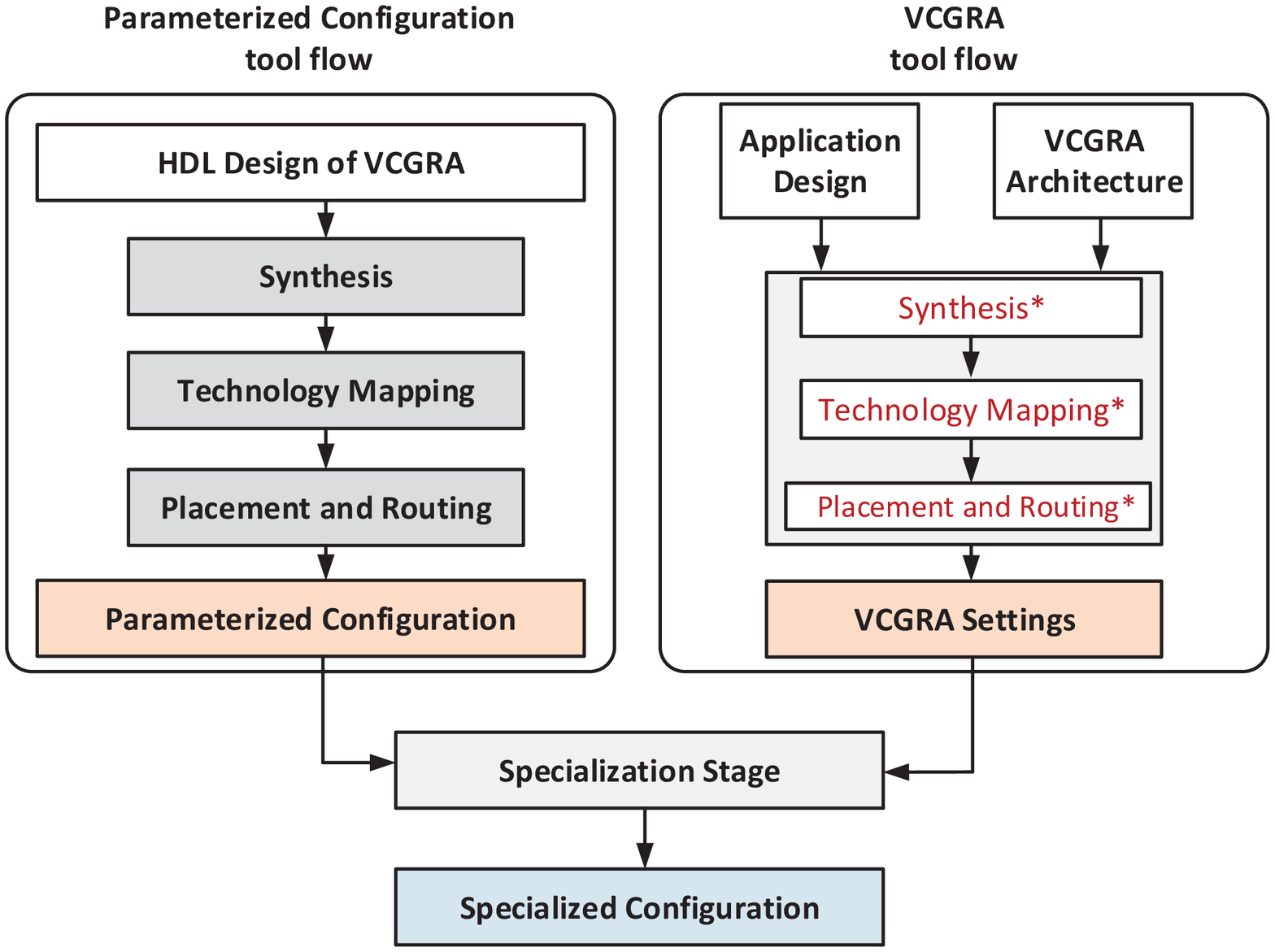}
\caption[caption]{Implementation of applications on a VCGRA using the parameterized configuration tool flow. \\\hspace{\textwidth} Note: * indicates steps considering PEs as a basic programmable component.}
\label{fig:vcgra_toolflow}\vspace*{-0.5cm}
\end{figure}

A fully parameterized VCGRA implementation is explained by the authors in~\cite{amit_5}. 
The tool flow to implement a fully parameterized VCGRA is depicted in Figure~\ref{fig:vcgra_toolflow}.
The VCGRA tool flow makes use of a VCGRA architecture that defines the granularity as well as the possible functionality of the PEs and describes the possible ways the PEs are interconnected. 
The implementations of the PEs and VSBs are performed in the parameterized reconfiguration tool flow (left hand side of Figure~\ref{fig:vcgra_toolflow}).

In the VCGRA tool flow (right hand side of Figure~\ref{fig:vcgra_toolflow}), the user will determine the VCGRA settings that will configure the required configurable components of the VCGRA to realize the desired application. 

The higher level VCGRA tool flow that produces these VCGRA settings consists of a synthesis and a mapping tool in which the textual description of the application design is parsed and converted into a netlist of Processing Elements (PEs). 
Next, we perform placement of the synthesized netlist of PEs on to the virtual PEs of the VCGRA architecture. 
The tool flow has to take care that all the inter- and intra-connect links of the VCGRA are implemented on the VCGRA architecture's communication network. 
We make use of a router to establish optimal connections between the placed elements of the VCGRA architecture. 
The Place and Route (PaR) result determines what the functionality of each PE is and how the communication network is exactly used and that this is reflected in the VCGRA settings.

Because the basic programmable element in the VCGRA tool flow is a PE, the tools (synthesis, mapper, place and route) need considerably less complexity and time to generate the settings values than the standard FPGA compilation would. 
This is because the higher abstraction level reduces the problem size and therefore the tools are faster. 
If the application design specification changes with the same VCGRA platform then we can generate the settings values much faster than processing the new design with the standard FPGA tool flow. 

Using the parameterized reconfiguration flow gives a set of parameterized VCGRA components. 
The settings values are then combined with these parameterized components in the specialization stage and this results in the final reconfiguration bitstreams automatically. 
A detailed explanation on parameterized configuration tool flow is presented in~\cite{karel_vcgra}.

In a conventional VCGRA implementation, the settings registers drive a generic design that can handle all different implementation possibilities and they are updated using a dedicated bus. 
However, in the parameterized VCGRA implementation, the settings registers of each PE and the routing switches are updated by reconfiguring each frame of the FPGA that contains setting bits of the VCGRA, thus optimizing the PE ad SB to a dedicated function. 
This is usually accomplished by read-modify and write back frames of the FPGA (\emph{micro-reconfiguration}).

For a VCGRA application that contains dynamic Network-On-Chips or PEs that require cycle-by-cycle context switching, we cannot afford the cost of reconfiguring so often and therefore, such applications may not be suitable to be handled by a parameterized VCGRA.

However, in the case of much less frequent reconfiguration needs, the parameterized reconfiguration reduces the overhead of the conventional VCGRA as follows:
\begin{itemize}
\item The settings registers of the VCGRA are mapped on the configuration memory and therefore, the need of a dedicated bus to update the settings register is avoided nor do we need to reserve application memory to implement the settings registers. 

\item The PEs of the VCGRA are optimized by symbolic constant propagation that is integrated within the parameterized configuration tool flow.

\item Each VCGRA intra- and inter-connection is mapped onto lower level reconfigurable routing switches (TCONs). Therefore, we reduce the utilization of the LUTs for implementing the connection network.
\end{itemize}


\section{The heterogeneous VCGRA grid} \label{sec:vcgra_grid}
In~\cite{karel_vcgra} a specific VCGRA for regular expression matching is introduced, while the authors in~\cite{amit_5} outline a floating point MAC operator, which is specialized for image processing tasks. 
In this paper we describe a more generic VCGRA, which includes basic processing elements and flexible virtual Channels (VC). 
In comparison to the MAC-version described in \cite{amit_5}, users are not restricted to MAC-operations and the application data-flow graphs can easily be mapped onto the VCGRA.
Thus, it is not necessary to massively modify the application for acceleration. Having a more general VCGRA offers another level of abstraction where the user can evaluate the suitability of a VCGRA for different kinds of applications. 
In addition, as all processing elements are working in parallel, the processing of the data can be pipelined as well as the proposed grid can be optimized for a specific application class, which results in higher throughput. 
Our design is currently focused on task graph representations of data-flow-oriented applications.
However, we plan to extend the functionality of our hardware to also support designs which contain more control-flow oriented code. 
For this reason the extension of the graph representation with control flow operations is planned. 

The methodology is programming language independent, because currently the toolchain's input is the data-flow graph of an application. 
Nodes of a graph represent the processing element functions, while edges show the dependencies and the dataflow between the processing elements. 
Currently we support arithmetic operations (addition, subtraction, multiplication, division) as well as comparison (greater than, equal to). 
In addition a PE also has modes for buffering a value and support for an idle state. 
Buffering is necessary for resolving data dependencies between node inputs from different levels.
\begin{figure}
 \centering
 \includegraphics[width=7cm]{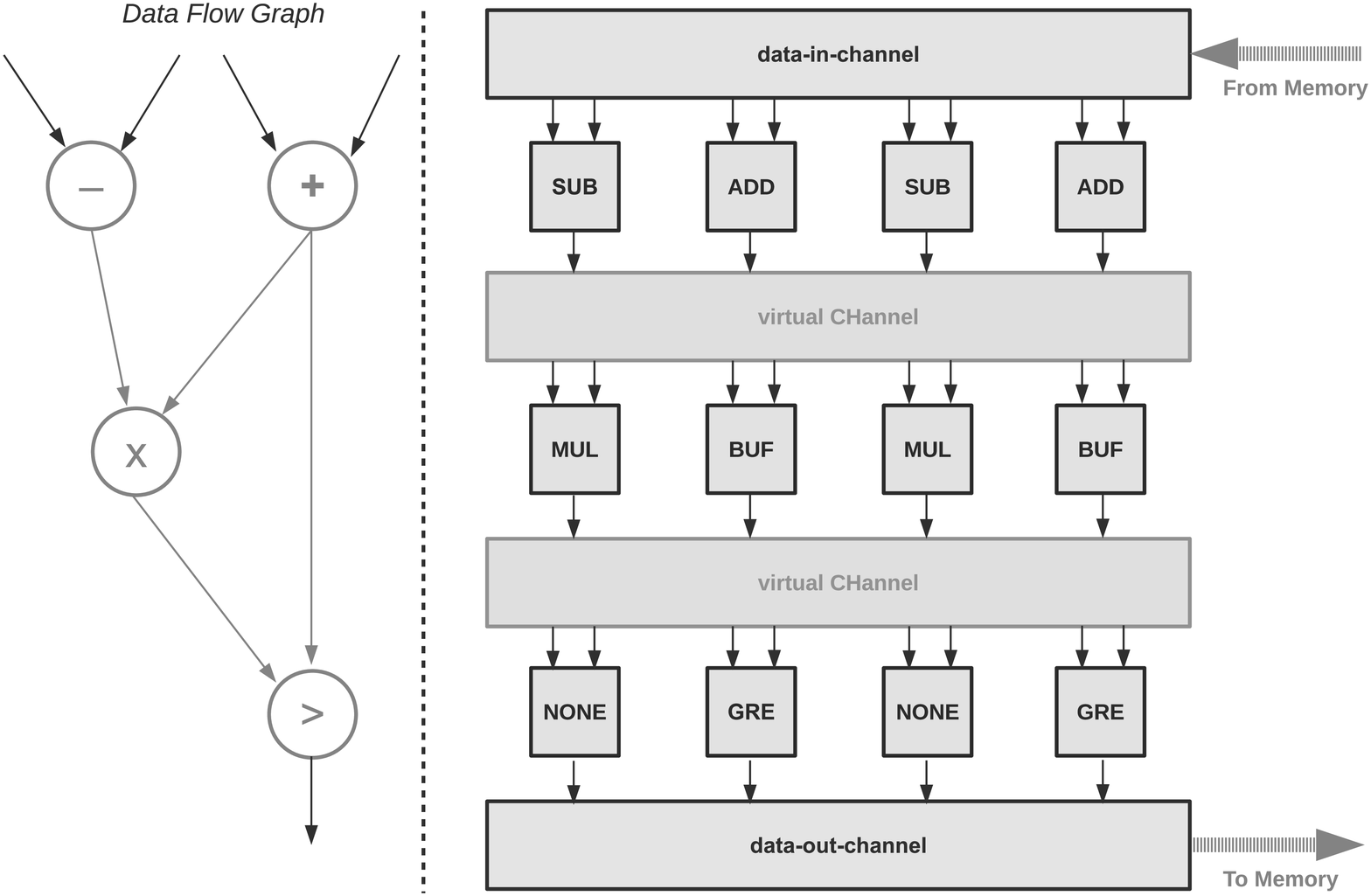}
 \caption{An overview of design of parameterized VCGRA.}
 \label{fig:vcgra_overview} 
\end{figure}
Data flow is oriented from the top to the bottom, which is depicted in a simple example in Figure \ref{fig:vcgra_overview}.

The grid of processing elements is organized in levels, whereby two levels are divided by one intermediate virtual channel.
This kind of design was chosen to enable the use of pipelining in the architecture. 
Every level of processing elements works as a pipeline stage. 
A specific kind of a VC is used as memory interface. 
It connects the grid to a microprocessor, which controls the VCGRA execution using any desired communication interface. 
The distribution of incoming data to the first row of processing elements is controlled by an external configuration signal. 
A synchronous start signal enables the execution in the first level of processing elements, when all incoming data dependencies are fulfilled. 
The start signal is to be controlled from outside the VCGRA, the other levels of PEs are synchronized with their predecessors within the array.
No additional control from a microcontroller is necessary. 
When the input data has been processed by the VCGRA, the processing system is notified to fetch the output data. 
The operation of the processing elements as well as the routing within the VC is realized by reconfiguration using the TLUT/TCON tool flow. 
The example in Figure \ref{fig:vcgra_overview} also shows opportunities for acceleration. 
If the grid is big enough, multiple instances of the same graph can be implemented.

We created a tool that eases the task of designing VCGRAs with different shapes. 
In addition to the rectangular style, where every row contains the same number of PEs we support an arbitrary number of inputs and outputs at a VC which leads to application specific grid designs if necessary. 
Automatic generation of these grids for a specific application class is currently work in progress. 
The functionality of the processing elements is extendable. 
For instance, we also experimented with PEs enabling floating point operations for addition and multiplication. 
The currently used PE structure has been optimized when compared to the MAC-operator presented in \cite{amit_5}.

\begin{figure}
\centering
\includegraphics[width=7cm]{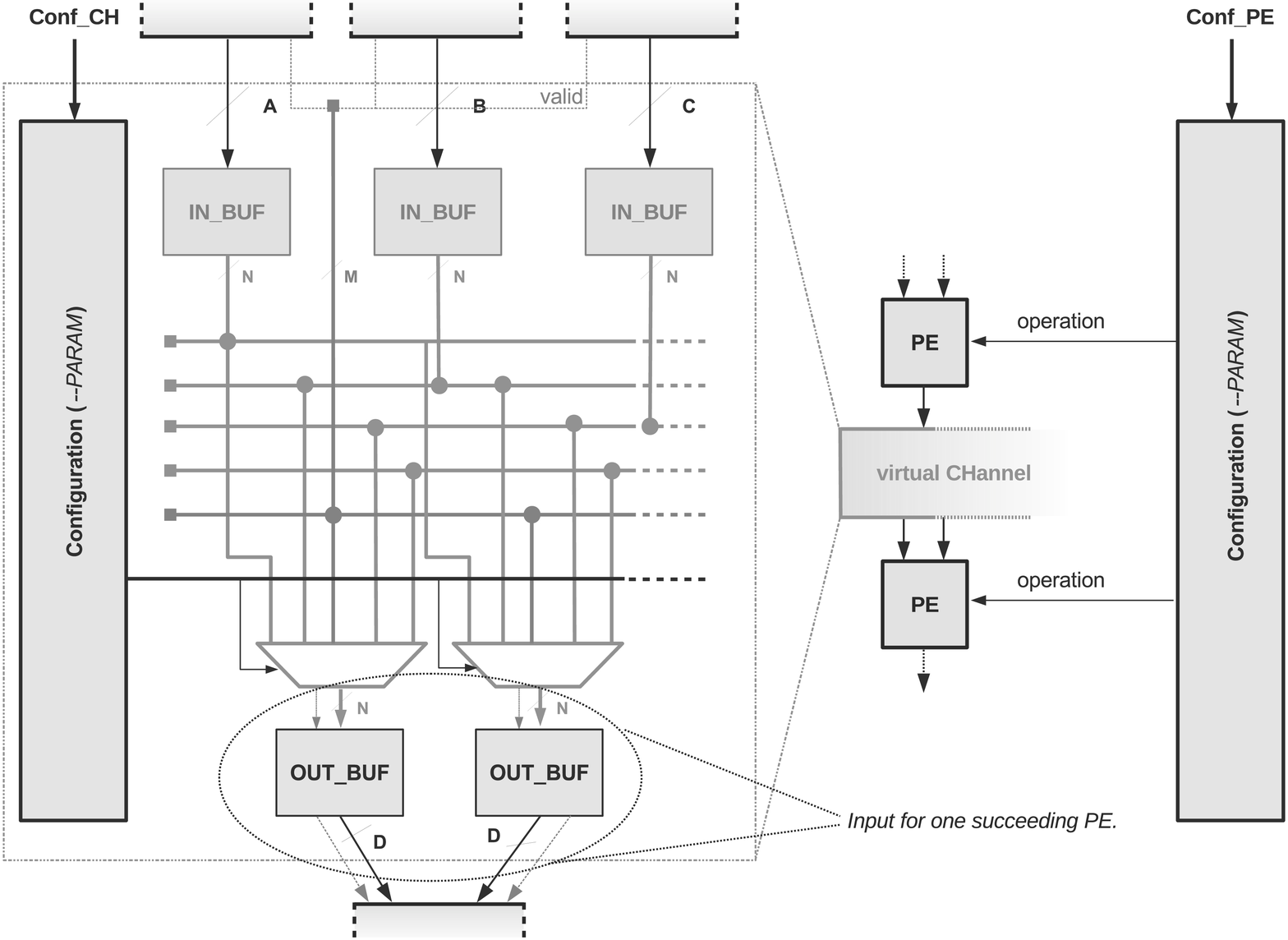}
\caption{Details of the proposed parameterized VCGRA.}
\label{fig:switch}
\end{figure}

\subsection{A fully Parameterized Processing Element (PE)}
The PE is designed as a finite state machine containing three stages: \emph{AWAIT}\_\emph{DATA}. \emph{PROCESS}\_\emph{DATA}, \emph{VALID}\_\emph{DATA}. 
Normally, it performs an operation on it's two inputs, which is set by a parameterized configuration input. 
The result is saved in an output buffer and is set to be valid for one cycle. 
To synchronize the inputs of a PE the two inputs have to be enabled.
This is done by using the valid signal from a previous PE in an upper level.
However, incoming values are buffered in every clock cycle. 
The current grid with its PEs is designed to support pipelined data flow applications. 
Therefore, no temporary results are saved within a processing element. 
We experiment with a MAC operation and design an element with a buffer to save the accumulated result, but we do not support graph mapping for that operation yet.
The data bitwidths of input and output are configurable.
However, the bitwidths of the two input values of a PE have to be equal. 
As shown in Figure \ref{fig:switch}, the adjustment of the bitwidth is done within the virtual channels.

If a PE is used as a buffer, the previous VC links the same data to both inputs of a PE. 
Thus, both inputs are enabled by the same valid signal and the data is copied to the output of the PE.
If a processing element is unused in a configuration it is configured with \emph{NONE}. 
A VC can bind arbitrary data to a PE's inputs. 
With the \emph{NONE} configuration, the PE does not generate any output or change the valid signal to synchronize a successor.

The intra-connects of the PE are also parameterized and therefore, the reconfigurable connections within the PE are also mapped on the tunable connections (TCONs) using the TCONMAP mapper~\cite{karel_tcon}. 


\subsection{Parameterized Virtual Channel (VC)}
The architecture of a first version of a VC is shown in Figure~\ref{fig:switch}.
The implementation currently needs a lot of routing resources (specifically connection multiplexers).
However, as the design is specially suited to be implemented using the TLUT/TCON tool flow the huge amount of multiplexers and connections which are dependent on a parameterized input are expected to need a significantly reduced amount of resources in the implementation, compared to an implementation using vendor tools.

One multiplexer per output is used to connect one specified input with the configured output. 
The select-input line of a multiplexer handles the specialization and is set as a parameter for the TLUT/TCON tool chain. 
This allows the TCON tool flow to distinguish which connections are used mutually exclusive in time. 
As a result, these routing resources can be shared within the FPGA.

All inputs of the predecessors of a channel are buffered at the input of the channel.
The valid signals of all previous PEs are collected.
Every input of a succeeding PE has a multiplexer with as many inputs as predecessors of the channel and gets a signal vector of all validating signals. 
Depending on the configuration, the output multiplexer routes the data value and the corresponding validating signal to an output buffer of the channel. 
A channel input can be routed to several channel outputs. 
As symbolized with the different letters at the connections, the channel supports different bitwidths for data paths. 
The internal bitwidth is set to the biggest data input, which can occur within a configuration,
\begin{equation}
N = \text{max}\left\lbrace A,B,C,D,\cdots \right\rbrace
\end{equation}
while the bitwidth of the validation signal vector depends directly on the number of predecessors.
\begin{equation}
M = \#\text{predecessors}
\end{equation}
The bitwidth of the internal channel connections is known a priori during the analysis of the task graph and is currently not changeable. 
Moreover, the size of a multiplexer and its bitwidth (bw) of a configuration word depends on the number of inputs or predecessors and is also fixed during system generation.

\begin{equation}
\text{bw} = \lceil \log_2 \left\lbrace \# \text{predecessors} \right\rbrace \rceil
\end{equation}

Nevertheless, the usage of the TCON tool flow shows promising results, which are described in more detail in Section~\ref{sec:results}.

\subsection{Building a VCGRA}
The PE and VC are the basic elements of a VCGRA that provide flexibility regarding their functionality and data bitwidths.
Describing the whole VCGRA grid in VHDL is a time consuming task.
Therefore we developed a tool that automatically creates the VHDL top-level description of a VCGRA from a description of the hardware structure.
The only inputs needed are the number of input elements from memory and the structure of the grid.
The grid's structure is described by the number of processing elements in each level of the architecture and the elements' input and output bitwidths.
All other parameters (e.g. for the channels) are automatically derived from the mentioned input data.
The tool's output is VHDL code defining the hardware structure of the grid.


\vspace*{-0.3cm}

\section{Edge Detection }\label{sec:application}
\begin{figure}[!t]
\centering
\includegraphics[width=8cm]{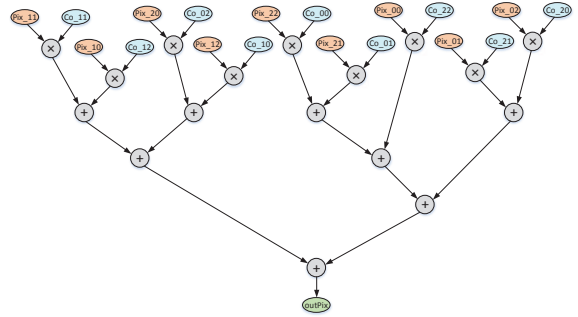}
\caption{Task graph representation of a 3 $\times$ 3 filter mask.}
\label{fig:taskgraph} 
\end{figure}

\begin{figure}[!t]
\centering
\includegraphics[width=7cm]{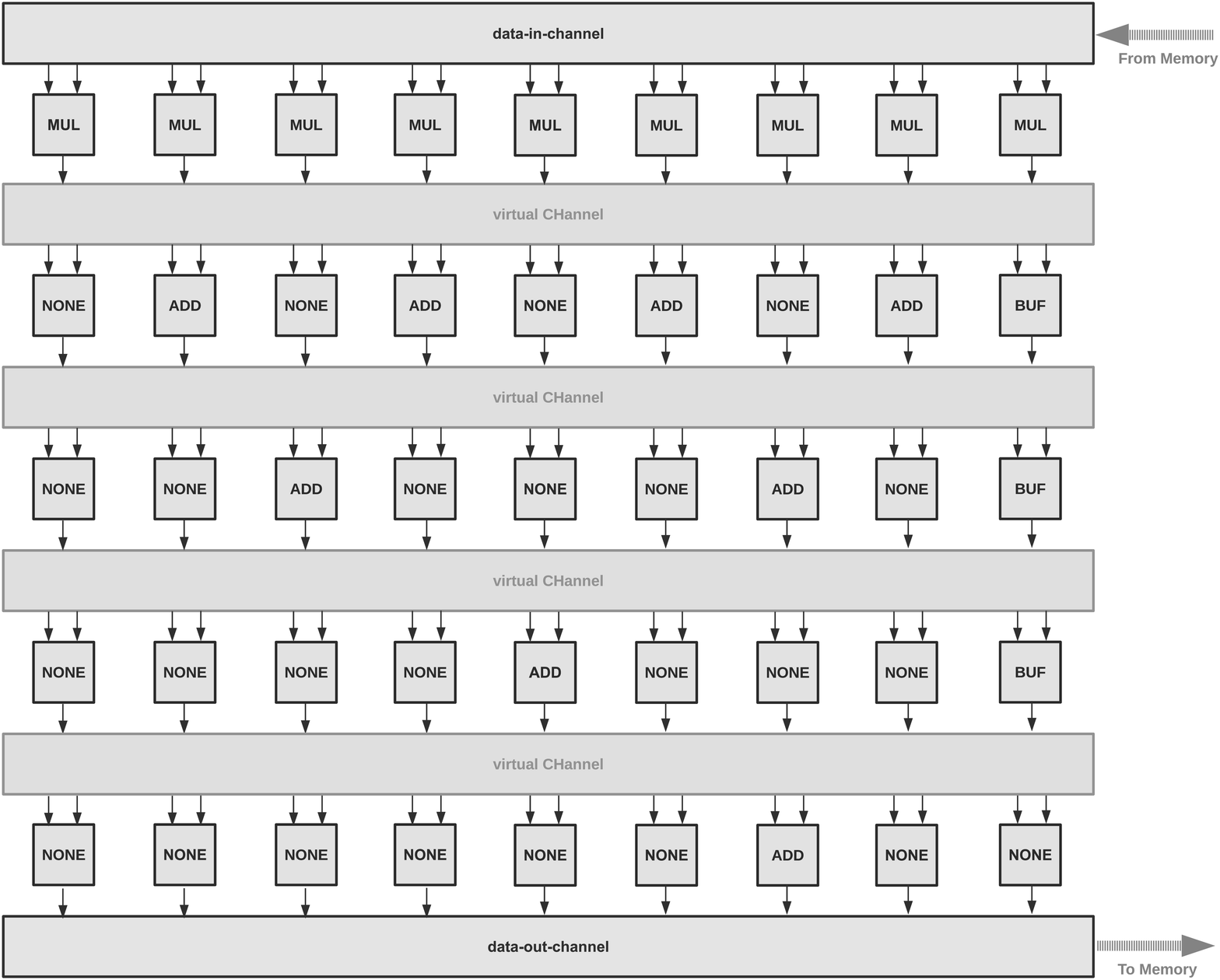}
\caption{VCGRA grid for the Sober edge detection filter.}
\label{fig:VCGRA_TG}
\end{figure}

For demonstration purposes we implemented the Sobel edge detection kernel on the proposed VCGRA. The Sobel filter is used for edge detection. An algorithm for Sobel edge detection filter is shown in Algorithm 1.

\begin{algorithm}
\footnotesize
\caption{Edge Detection}
\begin{algorithmic}[1]
\Procedure{Sobel}{$image$} \Comment{grayscale image}
\State $center \gets 0,0$ \Comment{setpoint of kernel}
\ForAll{pixel in image}
\State $pos \gets pixel\_coordinates$ \Comment{pixel position in image}
\State $sum \gets 0$
\For{$j \gets -1, 1$}
\For{$i \gets -1, 1$}
\State $temp \gets sobel[center + j][center  + i]$
\State $\times pixel[pos - j][pos - i]$
\State $sum \gets sum + temp$
\EndFor
\EndFor
\State $image[pos] \gets sum $
\EndFor
\EndProcedure
\label{alg:sobel}
\end{algorithmic}
\end{algorithm}




The setpoint of the Sobel kernel is set to the midpoint of the mask. 
Every pixel of an image is convolved with the kernel.
The result of the convolution is saved at the current position of the filter mask's setpoint in the image. 
A task graph representation of the algorithm is shown in Figure \ref{fig:taskgraph}. 
It shows the kernel code of the inmost loop. 
Blue nodes are pixel values, which lay underneath the kernel mask; red nodes are the corresponding filter coefficients. 
A gray node symbolizes an operation and is mapped to a corresponding PE. 
The edges are managed by the configuration of the virtual channels.
At least, the green node symbolizes a result of the convolution for a single pixel value.
We used a small image processing kernel for demonstration, because a task graph becomes very huge for bigger masks.
However, we are also able to implement bigger kernels on a VCGRA. 
The weighted pixel value of the multiplication on the right border of the array is buffered in every stage of the array until it is used in the last addition. 
The size of an array is arbitrary. 
For demonstration we choose an array which is as big as needed to implement all levels of the task graph.
However, it is also possible to choose bigger or smaller arrays.
For bigger arrays with more stages than necessary, an output value has to be buffered in every stage until it reaches the data output channel at the bottom. Bypassing of levels of the array is not supported. 

The VCGRA grid for the Sobel edge detection application is depicted in Figure~\ref{fig:VCGRA_TG}. The grid consists of 4 VCs and 45 PEs. For the simple implementation on a hypothetical FPGA, we have considered to design the rectangular VCGRA grid and hence we observe that the majority of the PEs are configured with the \emph{NONE} operation. However, this could be optimized by designing an inverted triangular grid.
\section{Results and Discussion}
\label{sec:results}
The VCGRA components described in the previous section were synthesized and were subjected to a Place and Route (P\&R) tool using the TPaR CAD tool~\cite{elias_tpar}. The P\&R was performed using the \emph{4LUT\_sanitized} FPGA architecture from VPR~\cite{vpr}. The results of P\&R for the VCGRA and its components are explained in the following subsections. 
\vspace*{-0.2cm}
\subsection{Virtual Channel (VC)}
The Virtual Channel is described in VHDL. The channel has parameterized connection multiplexers whose select lines are the parameter inputs. With the help of the TCONMAP mapper we were able to map the VC on the TCONs. Therefore, the major part of the VC do not need LUT to make them reconfigurable. 

\begin{table}
\renewcommand{\arraystretch}{0.8}
\caption{Resource utilization and P\&R results.}
\label{tab:par}
\centering
\resizebox{\columnwidth}{!}{
\begin{tabular}{|c|c|c|c|c|c|}
\hline
\bfseries \begin{tabular}[x]{@{}c@{}}  \end{tabular} & \bfseries \begin{tabular}[x]{@{}c@{}}LUTs (TLUTs) \end{tabular}  & \bfseries \begin{tabular}[x]{@{}c@{}}TCONs\end{tabular}  & \bfseries \begin{tabular}[x]{@{}c@{}}Logic\\Depth\\level\end{tabular}&
\bfseries \begin{tabular}[x]{@{}c@{}}WL\end{tabular}  & 
\bfseries \begin{tabular}[x]{@{}c@{}}mCW\end{tabular}\\
\hline\hline
\begin{tabular}[x]{@{}c@{}}VC\\ Conventional\end{tabular} & 
\begin{tabular}[x]{@{}c@{}}$176 (0)$\end{tabular} & 
\begin{tabular}[x]{@{}c@{}}$0$\end{tabular}  & 
\begin{tabular}[x]{@{}c@{}}$2$\end{tabular} & 3186 & 7\\ \hline

\begin{tabular}[x]{@{}c@{}}VC\\Parameterized\end{tabular} & 
\begin{tabular}[x]{@{}c@{}}$32 (0) $\end{tabular} & 
\begin{tabular}[x]{@{}c@{}}$72$\end{tabular}  & 
\begin{tabular}[x]{@{}c@{}}$1$\end{tabular} & 782 & 4\\ \hline

\begin{tabular}[x]{@{}c@{}} PE\\Conventional\end{tabular} & 
\begin{tabular}[x]{@{}c@{}}$408 (0)$\end{tabular} & 
\begin{tabular}[x]{@{}c@{}}$0$\end{tabular}  & 
\begin{tabular}[x]{@{}c@{}}$47$\end{tabular} & 3832 & 8\\ \hline

\begin{tabular}[x]{@{}c@{}}PE\\Fully\\Parameterized\end{tabular} & 
\begin{tabular}[x]{@{}c@{}}$387 (32) $\end{tabular} & 
\begin{tabular}[x]{@{}c@{}}$22$\end{tabular}  & 
\begin{tabular}[x]{@{}c@{}}$47$\end{tabular} & 3769 & 8\\ \hline

\begin{tabular}[x]{@{}c@{}}PE\_FP\\Conventional\end{tabular} & 
\begin{tabular}[x]{@{}c@{}}$2191 (0)$\end{tabular} & 
\begin{tabular}[x]{@{}c@{}}$0$\end{tabular}  & 
\begin{tabular}[x]{@{}c@{}}$47$\end{tabular} & 23388 & 10\\ \hline

\begin{tabular}[x]{@{}c@{}}PE\_FP\\Fully\\Parameterized\end{tabular} & 
\begin{tabular}[x]{@{}c@{}}$1668 (584) $\end{tabular} & 
\begin{tabular}[x]{@{}c@{}}$798$\end{tabular}  & 
\begin{tabular}[x]{@{}c@{}}$47$\end{tabular} & 17676 & 10\\ \hline

\begin{tabular}[x]{@{}c@{}}Grid\\Conventional\end{tabular} & 
\begin{tabular}[x]{@{}c@{}}$17066 (0)$\end{tabular} & 
\begin{tabular}[x]{@{}c@{}}$0$\end{tabular}  & 
\begin{tabular}[x]{@{}c@{}}$155$\end{tabular} & 176200 & 14\\ \hline

\begin{tabular}[x]{@{}c@{}}Grid\\Fully\\Parameterized\end{tabular} & 
\begin{tabular}[x]{@{}c@{}}$16099 (976) $\end{tabular} & 
\begin{tabular}[x]{@{}c@{}}$561$\end{tabular}  & 
\begin{tabular}[x]{@{}c@{}}$153$\end{tabular} & 169560 & 12\\ \hline

\end{tabular}
}\vspace*{-0.6cm}
\end{table}

The P\&R results of the VC implementation are tabulated in Table~\ref{tab:par}. From the top two lines of Table we observe 82\% of the logic are mapped on the reconfigurable physical switches (TCONs) instead of physical LUTs and multiplexers (as per the conventional implementation). We also observe a significant decrease of 76\% in wire length (WL) between conventional and parameterized implementation due to the fact that the minimum channel width (mCW) is reduced by 42\%. This optimization can be achieved at the cost of a reconfiguration time of 4.6 ms (not shown in Table).
\vspace*{-0.2cm}
\subsection{Processing Element (PE)}
We have designed a Processing Element that comes with two different versions: a fixed point PE and a floating point PE. The P\&R results of the fixed point and floating point PE are tabulated in Table~\ref{tab:par}.



The logic resources (LUTs) used by the fixed point PE are optimized by 5\% and we also observe a difference in wire length by 2\%. This optimization can be achieved  by investing a reconfiguration time costs of 3.4 ms. The PE contains 13\% of its design (TLUTs + TCONs) that are responsible to form a reconfigurable processing element. 

The floating point PE was built using an open source floating point library called \quotes{FloPoCo}~\cite{flopoco}. We used the FloPoCo floating point format with a 6-bit exponent and a 26-bit mantissa. We have not used any dedicated multipliers or adders while generating the floating point operators using the \quotes{FloPoCo} library. The floating point PE implementation was optimized by 24\% and a decrease in wire length by 25\% is also observed. This optimization can be achieved at the cost of a reconfiguration time of 88.5 ms. There is no difference in logic depth level and minimum channel width in both types of PEs. The PE contains 82\% of its design (TLUTs +TCONs) that are responsible to form the reconfigurable part of the processing element. The proposed floating point PE consumes 13\% less resources compared to a MAC operator presented in~\cite{amit_5}.


\vspace*{-0.2cm}
\subsection{A fully parameterized 4$\times$4 VCGRA grid}
A fully parameterized 4$\times$4 VCGRA grid was implemented using fixed point PEs and VCs. The P\&R results are tabulated in Table~\ref{tab:par}. The logic resources of the whole grid are optimized by 6\% and the wire length is reduced by 4\% due to a reduction in logic depth level by 2 and in minimum channel width by 2 as well. This optimization can be achieved at the cost of a reconfiguration time of 98.5 ms.


\vspace*{-0.2cm}
\subsection{Sobel filter}
To implement the Sobel filter we need 45 PEs and 4 VCs. To reconfigure all the processing elements and virtual channels it costs 156 ms and 18.4 ms of reconfiguration time respectively.\vspace*{-0.2cm}
\subsection{Compilation time}
The time taken to map the Sobel edge detection application is less than one second. The time taken to compile the hardware description of the VCGRA grid into bitstreams is approx. 1200 seconds. In the conventional implementation for every new image processing application, the development time would cost more than 1200 seconds. However, with the VCGRA approach, the total time to set up a new image processing application is very minimal since it cost only the mapping time and the total reconfiguration.

\vspace*{-0.1cm}

\section{Conclusion} \label{sec:conclusion}
We proposed a low cost heterogeneous VCGRA grid for image processing applications. The grid was implemented using the parameterized configuration technique. The proposed grid can be used as overlay architecture on a low cost FPGA platform that does not consist of hard coded primitives such as DSP blocks. As part of a demonstration we built the Sobel edge detection filter and the results show a promising improvement in the compilation times and thus bridging the gap between the application and the FPGA fabric. In the future we will build a parallel memory architecture with a dedicated reconfiguration bus for the VCGRA that will accelerate the reconfiguration speed and thus further reduce the reconfiguration time costs.

\section*{Acknowledgment}
This work was supported by the European 
Commission in the context of the H2020 FETHPC EXTRA project (\#671653). 
Special thanks to the HiPEAC (\#687698) collaboration grant for supporting a part of this research work.
\vspace*{-0.15cm}

\bibliography{Bibliography}

\newcommand{\specialcell}[2][c]{%
  \begin{tabular}[#1]{@{}c@{}}#2\end{tabular}}
\end{document}